\documentclass[a4paper,twocolumn,prl,nofootinbib,floatfix,showpacs]{revtex4}

\usepackage{amsmath}
\usepackage{amssymb} 
\usepackage{epstopdf}
\usepackage{graphicx}

\usepackage[usenames,dvipsnames]{pstricks}
\usepackage{epsfig}
\usepackage{pst-grad} % For gradients
\usepackage{pst-plot} % For axes

\usepackage{ulem}
\newcommand{\bra}[1]{\langle #1 |} 
\newcommand{\ket}[1]{| #1 \rangle }

\newcommand{\half}{\frac{1}{2}}
\newcommand{\BigO}[1]{\ensuremath{\operatorname{O}\bigl(#1\bigr)}}

\usepackage{color}
\definecolor{cbl}{rgb}{0,0,1}                % bleu

\begin{document}

\title{Open quantum random walks: bi-stability on pure states\\ and ballistically induced diffusion.}

\author{Michel Bauer ${}^{\spadesuit~}$, Denis Bernard 
  ${}^{\clubsuit~}$ and Antoine Tilloy ${}^{\clubsuit~}$} 
  \email{michel.bauer@cea.fr, denis.bernard@ens.fr, antoine.tilloy@ens.fr}.

\date{\today} 

\affiliation{ ${}^\spadesuit$ Institut de Physique Th\'eorique de Saclay,
%\footnote{CEA/DSM/IPhT, Unit\'e de recherche associ\'ee au CNRS}, 
  CEA-Saclay $\&$ CNRS, 91191 Gif-sur-Yvette, France.\\
  $^\clubsuit$
  %Laboratoire de Physique Th\'eorique de l'Ecole Normale Sup\'erieure, \\
  Laboratoire de Physique Th\'eorique de l'ENS, 
  CNRS $\&$ Ecole Normale Sup\'erieure de Paris, France }

\preprint{ Preprint IPhT 2013/??? ; arxiv?/??? }
        
\begin{abstract}
Open quantum random walks (OQRWs) deal with quantum random motions on the line for systems with internal and orbital degrees of freedom. The internal system behaves as a quantum random gyroscope coding for the direction of the orbital moves. We reveal the existence of a transition, depending on OQRW moduli, in the internal system behaviors from simple oscillations to random flips between two unstable pure states. This induces a transition in the orbital motions from usual diffusion to ballistically induced diffusion with large mean free path and large effective diffusion constant at large time. We also show that mixed states of the internal system are converted into random pure states during the process. We touch upon possible experimental realizations.
\end{abstract}


\maketitle

\section{Introduction}
Random walks \cite{feller} are ubiquitous in our understanding of physical phenomena with plethora of applications in biology or economics. They are instrumental in mathematics and computer science. Quantum generalizations have been considered  decades ago \cite{Qwalk_review} and they find numerous applications in quantum computation or quantum cryptography \cite{Qcomp}. They have recently been experimentally implemented \cite{meschede, Peruzzo, Schmitz}. Drastically influenced by quantum interferences, quantum random walks behave very differently from their classical analogues, for instance they do not diffuse in the same way \cite{Qwalk_review, Konno-Gosw}.  

Open quantum random walks (OQRWs) were introduced \cite{attal_etc} using concepts from quantum dynamical maps \cite{Qdynamics} aiming at incorporating decoherence effects. They specify random motions of quantum systems with both internal and orbital degrees of freedom (d.o.f), and these moves depend on interactions with quantum coins. Contrary to quantum random walks, OQRWs implement resettings of the quantum coins at each time step, and this difference has profound consequences. 

Studying classes of OQRWs we find a transition in their behaviors separating usual diffusions from ballistically induced diffusions with large mean free path between trajectory flips. Of course diffusion is always due to ballistic behaviors at a small enough scale, what matters is the time separation between flips. In OQRWs, these are not due to disordered collisions but to abrupt tilts of the internal gyroscope induced by the interaction with quantum coins and their measurements. Behaviors in the ballistic regime are consequences of random switches of the internal state between unstable pure states. 

OQRW is a too recent research field to reliably predict its future domains of application which, we may expect, will include quantum deformation of that of classical random walks. The scaling limit we discuss here provides an elementary and pathology free definition of quantum Brownian motion \cite{Caldeira-Leggett, Froehlich} with clear potential outputs to this subject \cite{BBT}. One may also contemplate applications of the mechanism of ballistically induced diffusion, and its large effective diffusion constant, to possibly quantum mechanically induced biological phenomena, especially photosynthetic energy transfer \cite{Engel,Mohseni}. Notice that our results about convergence to unstable pure states apply to a Qbit interacting repeatedly with series of Qbits without considering orbital d.o.f.'s.

\section{Open Quantum Random Walks and their continuous limit} 
\subsection{Definition}
To be closer to possible experimental realizations and to quantum trajectory theories \cite{Qtrajectory,Qtraj2}, we define OQRWs using a picture slightly different but equivalent to \cite{attal_etc} in which the system interacts recursively with identical quantum coins, called probes \cite{Qnoise}. We shall represent the quantum system, with Hilbert space ${\cal H}_c\otimes {\cal H}_o$, as a particle with internal and orbital d.o.f.'s: the former may be represented by an effective spin or by colors and the latter, labeled as $\ket{n}_o$, either refer to localized positions on the line or to energy levels in a potential well. The probe Hilbert space ${\cal H}_p$ is chosen to be two dimensional with a specified basis $\{\ket{\pm}_p\}$. At each time step, the system interacts quantum mechanically with one sample of identically prepared copies of the probe on which a measurement is performed after the interaction period. The system-probe interaction is such that if the out-going probe is measured in the state $\ket{+}_p$ (resp. $\ket{-}_p$) the system moves by one step to the right (resp. to the left) along the line, and this move is accompanied by a modification of the internal d.o.f's. The system position is thus slave to the measurement out-puts.

Although experimental realizations of OQRWs do not yet exist we may contemplate possible scenarios. One may imagine using ions trapped in harmonic potentials, as in \cite{wineland,cirac}, each ion being possibly in two states with different angular momenta, and photons as probes. For an appropriately adjusted frequency and linearly polarized in-going photons, the ion-photon interaction may induce internal flips and energy shifts conditioned on the measurements of out-going photons \cite{problem}. One may also imagine using cold atoms with internal d.o.f's and localized on potential lattices, as in \cite{meschede}, and probing them coherently with photons \cite{problem}. If one is only interested in the internal system \cite{classical_slave}, a set-up dealing with recursive couplings of a Qbit to series of probe Qbits, as in \cite{blatt}, may be considered.

To make this description concrete, let ${\cal H}_{\rm sys}:={\cal H}_c\otimes {\cal H}_o$ be the system  Hilbert space, with ${\cal H}_c$ and ${\cal H}_o$ respectively associated to the internal and orbital d.o.f's. We take ${\cal H}_c$ finite dimensional and ${\cal H}_o\simeq \mathbb{C}^{\mathbb{Z}}$ with  orthonormal basis $\{\ket{n}_o,\, n\in\mathbb{Z}\}$. Let $U$ be the unitary operator acting on ${\cal H}_{\rm sys}\otimes{\cal H}_p$ coding for the system-probe interaction. We demand that its action on states $ \ket{\psi}_c\otimes\ket{n}_o\otimes\ket{\phi}_p$ gives the entangled normalized states:
\[(B_+\ket{\psi}_c)\otimes\ket{n+1}_o\otimes\ket{+}_p +(B_-\ket{\psi}_c)\otimes\ket{n-1}_o\otimes\ket{-}_p\]
for any $\ket{\psi}_c\in{\cal H}_c$. Unitarity imposes  $B_+^\dag B_++ B_-^\dag B_-=\mathbb{I}$.

OQRWs consist in iterating system-probe interactions and out-going probe measurements. Since the latter are random with probabilities governed by quantum mechanics, this yields stochastic evolutions called quantum trajectories \cite{Qtrajectory,Qtraj2}. If the system density matrix is initially localized in the orbital space, say $\rho_0\otimes \ket{x_0}_o\bra{x_0}$, it remains so after each iteration with internal density matrix $\rho_n$ and orbital position $x_n$. These are randomly updated, 
\begin{equation} \label{eq:dilatOQRW}
\rho_n\otimes\ket{x_n}_o\bra{x_n}\to \frac{B_\pm \rho_n B_\pm^\dag}{p^\pm_n}\otimes \ket{x_n\pm1}_o\bra{x_n\pm1},
\end{equation}
with probability $p^\pm_n:={\rm tr}_{{\cal H}_c}(B_\pm \rho_n B_\pm^\dag)$. The process $n\to (\rho_n,x_n)$ is Markovian on a probability space whose events are the recursive out-put probe measurements. By construction the mean system density matrix evolves according to the OQRW quantum dynamical map \cite{attal_etc}, and the mean internal density matrix $\bar\rho_n:=\mathbb{E}[\rho_n]$ satisfies $\bar\rho_{n+1}=B_+ \bar\rho_n B_+^\dag+B_-\bar\rho_n B_-^\dag$. In absence of internal d.o.f's OQRW behaviors parallel those of classical random walks. We take ${\cal H}_c\simeq \mathbb{C}^2$ and represent the internal system by an effective spin one-half gyroscope.

\begin{figure}
\mbox{\includegraphics[width=0.435\textwidth]{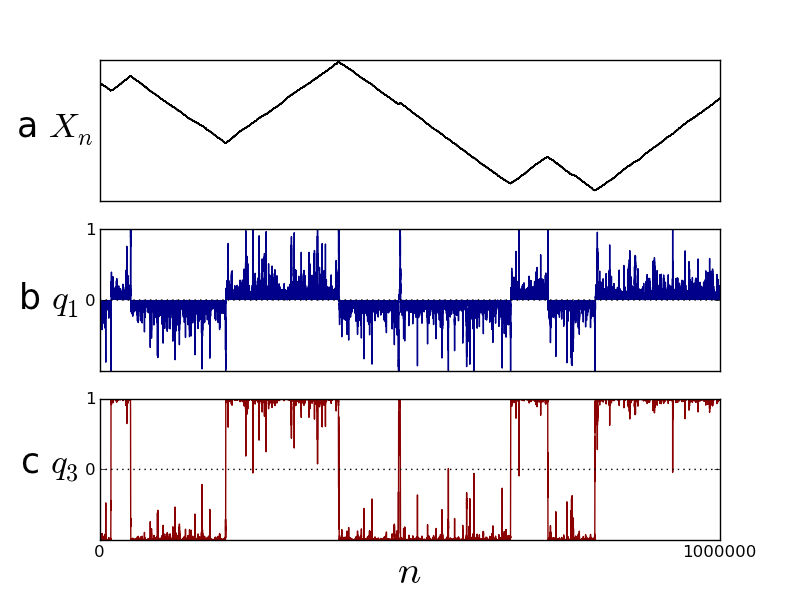}}
\caption{(Color online) Typical OQRW trajectory generated by $B_\pm$ as in the text with $u=1.1$, $v=1.00$ and $r=-s=0.00015$ (corresponding to $a^2/\omega_0 \simeq 4$ at the continuous limit): (a) position $X_n$, (b)$\&$(c) $\sigma^1\,\&\,\sigma^3$-components of $\rho_n$.}
\label{fig:discrete_ballistic}
\end{figure}

\subsection{Heuristics}
%\sout{Solutions of the unitary constraint may be parametrized as $B_\pm=U_\pm M_\pm$ with $U_\pm$ unitary and $M_\pm$ hermitian with $M_+^2+M_-^2=\mathbb{I}$. These are the OQRW moduli.} 
In the numerical simulations we look for OQRWs generated by matrices of the form: \[B_+=\delta^{-1}\big(\begin{smallmatrix} u&r\\s&v\end{smallmatrix}\big), \;
B_-=\delta^{-1}\big(\begin{smallmatrix}-v&s\\r&-u\end{smallmatrix}\big)\] 
with $\delta=\sqrt{u^2+v^2+r^2+s^2}$. This is not the most general parametrization but we use it only to give numerical illustrations of our results which concern mostly the continuous limit. In the scaling limit the most general matrices solutions of the unitarity constraint and consistent with the existence of a continuous limit will be:
\[B_\pm=\frac{1}{\sqrt{2}}[\mathbb{I}\pm\sqrt{\epsilon} N + \epsilon(-iH_\pm\pm M -\frac{1}{2} N^\dag N)+o(\epsilon)] \]
with $\epsilon$ a small parameter and $H_\pm$, $M$ Hermitian but not $N$. 
We take $H:=\half(H_++H_-)=\omega_0\, \sigma^2$ and $N=a\, \sigma^3$ with $\sigma^{1,2,3}$ the usual Pauli matrices. Numerical simulations are done with real matrices $B_\pm$, and these fit with our choice of $H$ and $N$. We fix $r=-s$ but vary $u$ and $v$, and this amounts to fix $\omega_0$ but modify $a$.

Numerical simulations reveal the existence of different regimes for OQRWs corresponding, in the scaling limit to $a^2/\omega_0$ below or above a critical value. For $a^2/\omega_0$ small enough, the position $x_n$ is nearly Brownian and the internal density matrix $\rho_n$ oscillates almost regularly, see Fig.\ref{fig:discrete_oscillant}. More interesting behaviors occur for $a^2/\omega_0$ above the critical value, see Fig.\ref{fig:discrete_ballistic}. The position $x_n$ follows a random seesaw trajectory, with tiny fluctuations, whose slopes are determined by the internal state which fluctuates around two unstable fixed points and toggles randomly from one to the other. The abrupt changes in the position moves are due to the random flips of the internal gyroscope. The parameter $a^2/\omega_0$ controls the mean free path between flips. Although ballistic on this time scale, the position is diffusive on larger time scale. Albeit being not completely obvious to prove, this result is expected from the 
central limit theorem. In addition, whatever the initial value, the internal density matrix converges rapidly to pure states, so that the fixed points are also pure states. It is quite remarkable that series of indirect probe measurements project mixed states on pure states. The progressive collapses elegantly observed in \cite{Haroche}, and proved in \cite{BB11}, is a particular illustration of this phenomena, but in OQRW context the target states keep on evolving randomly.

\begin{figure}
\mbox{\includegraphics[width=0.5\textwidth]{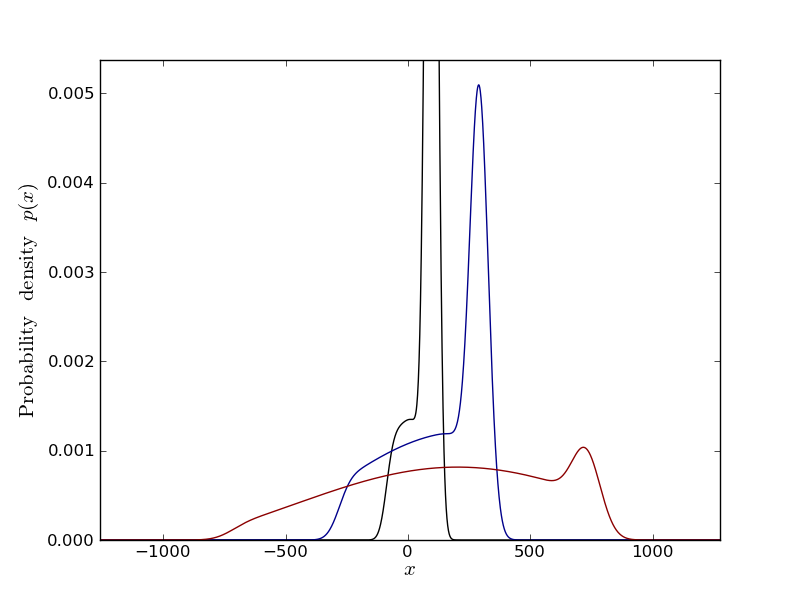}}
\caption{(Color online) Probability distribution $p(x,T)$ for a OQRW generated by $B_\pm$ as in the text with $u=1.1$, $v=1.00$ and $r=-s=0.0006$ (corresponding to $a^2/\omega_0 \simeq 2$ at the continuous limit). In black (narrowest curve) $T=2000$, in blue (middle curve) $T=6000$, in red (widest curve) $T=15000$. The right moving peak corresponds to the trajectories with no gyroscope flip, i.e. that went up in a quasi straight line. The nearly uniform plateau in the middle is the signature of the trajectories with one flip. More details can be found in the appendix. Notice that for large time the profile indeed starts to become Gaussian.}
\label{fig:pdf}
\end{figure}

\begin{figure}
\mbox{\includegraphics[width=0.5\textwidth]{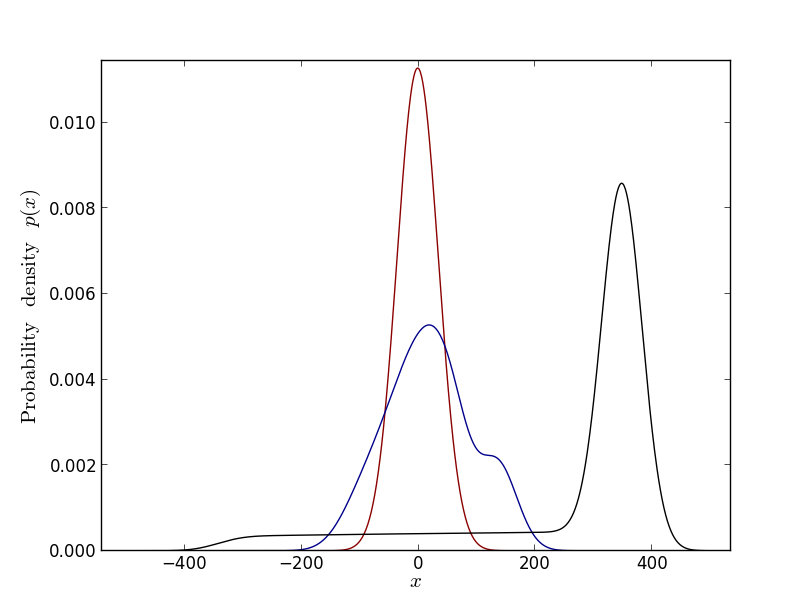}}
\caption{(Color online) Probability distribution $p(x,T)$ at fixed time $T=5000$ for a OQRW generated by $B_\pm$ as in the text with $v=1.00$, $r=-s=0.0006$ and $u=1.005$ for the narrowest distribution in red (corresponding to $a^2/\omega_0 \simeq 0.1$ at the continuous limit), $u=1.05$ ($a^2/\omega_0 \simeq 1$)  for the distribution with the medium width in blue and $u=1.15$ ($a^2/\omega_0 \simeq 3$) for the widest distribution in black. As expected, for small $a^2/\omega_0 \ll1$, the distribution looks Gaussian and gets more and more skewed as this ratio increases.}
\label{fig:pdfEqualTime}
\end{figure}

This peculiar behavior bears similarities with that of a noisy particle in a double well potential subject to Kramer's transitions from one well to the other. This is the picture that we are going to make explicit in the following. In that case we also look at a more common observable (see Figs.\ref{fig:pdf}$\&$\ref{fig:pdfEqualTime}), i.e. the probability distribution function (p.d.f.) of the process and notice an interesting intermediate time scale giving rise to a skewed profile which is a direct consequence of the seesaw profile of the trajectories. In the Appendix, we present a simple classical model which mimics this behavior.

\subsection{State purification}

The convergence towards pure states can be understood as follows. Let $\Delta_n:=\det\rho_n \geq 0$ be the determinant of the internal density matrix. In dimension 2, it vanishes only for pure states. A simple computation shows that $\mathbb{E}[\Delta_n^{1/2}]=c^n\, \Delta_0^{1/2}$ with $c:=\det^{\half}(B_+B_+^\dagger)+\det^{\half}(B_-B_-^\dagger)<1$ unless $B_+$ and $B_-$ are proportional to unitary matrices and the walk is classical which thus implies $\lim_{n\to\infty}\mathbb{E}[\Delta_n^{1/2}]=0$, the convergence being exponentially fast.

Actually we can prove that $\lim_{n\to\infty}\Delta_n^{1/2}=0$ almost surely using the sub-martingale convergence theorem of probability theory \cite{probas}. Indeed, computing the mean of $\Delta^{1/2}_{n+1}$ conditioned on the $n$-first out-put measurements gives $\mathbb{E}[\Delta_{n+1}^{1/2}|{\cal F}_n]=c\, \Delta_n^{1/2}<\Delta_n^{1/2}$, so that $\Delta_n^{1/2}$ is a sub-martingale, and since it is bounded, it converges almost surely and in $\mathbb{L}^1$. The limit can only be zero as the limit in $\mathbb{L}^1$ is zero and  the internal density matrix localizes on pure states.

\subsection{Fokker Planck picture}

In a continuous limit, the mean system density matrix reads $\int {\hskip -0.1 truecm} dx\, \rho(x,t)\otimes \ket{x}_o\bra{x}$ with $p(x,t):={\rm tr}_{{\cal H}_c}\rho(x,t)$ the probability density to find the system at position $x$ at time $t$, and $\bar \rho_t:=\int {\hskip -0.1 truecm} dx\, \rho(x,t)$ the mean internal state. At each time step $dt$, it is updated using OQRW rules (\ref{eq:dilatOQRW}), 
\[\rho(x,t+dt)= B_-\rho(x+dx,t)B_-^\dag+B_+\rho(x-dx,t)B_+^\dag.\]
A continuous limit exists if one imposes the scaling relation $\epsilon=dt=dx^2$ \cite{scaling}. Taylor expansion then gives:
\begin{eqnarray} \label{eq:FPrho}
\partial_t\rho= \half \partial_x^2\rho- (N\partial_x\rho+\partial_x\rho N^\dag) - i[H,\rho]+L_N(\rho),
\end{eqnarray}
with Lindbladian $L_N(\rho):=N\rho N^\dag - \half(N^\dag N\rho+ \rho N^\dag N)$. Eq.(\ref{eq:FPrho}) mixes pieces from diffusive Fokker-Planck equation and from Lindbladian quantum evolution for $\bar \rho_t$ \cite{lindblad}. The term $(N\partial_x\rho+\partial_x\rho N^\dag)$ is at the origin of the ballistic behavior seen in Fig.\ref{fig:discrete_ballistic} and of the large effective diffusion constant but the Hamiltonian term is required for the tilting effect. The probability density $p(x,t)$ is not associated to a Markov process and does not satisfy a linear equation but it becomes Gaussian at large $t$.

\begin{figure}
\mbox{\includegraphics[width=0.435\textwidth]{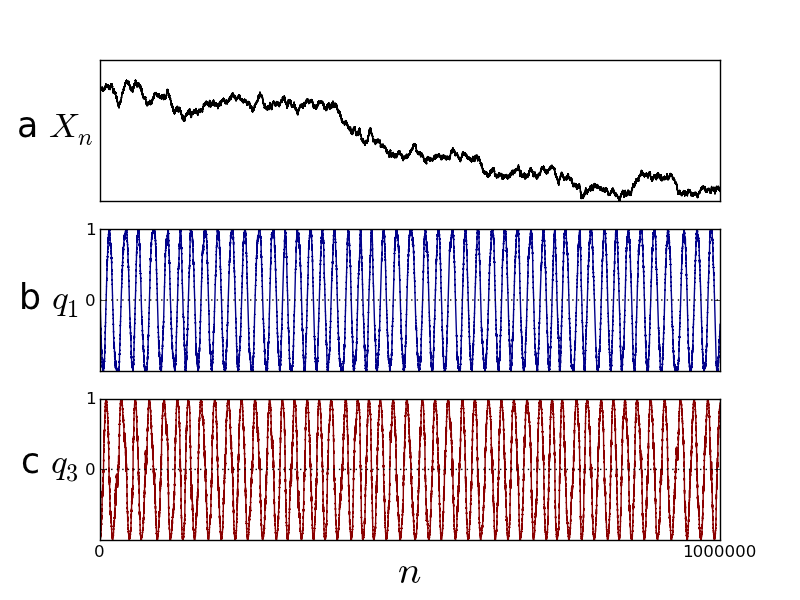}}
\caption{(Color online) Typical OQRW trajectory generated by $B_\pm$ as  in the text with $u=1.005$, $v=1.00$ and $r=-s=0.00015$ (corresponding to $a^2/\omega_0 \simeq 0.2$ at the continuous limit): (a) position $X_n$, (b)$\&$(c) $\sigma^1\,\&\,\sigma^3$-components of $\rho_n$.}
\label{fig:discrete_oscillant}
\end{figure}

\subsection{Quantum trajectory}
Let us now make precise the heuristic description by deriving the stochastic differential equations (SDEs) governing OQRWs in the scaling limit. OQRWs are defined on the probability space whose events are the series $(s_1,s_2,\cdots)$ with $s_k=\pm$ depending whether the $k$-th out-going probe is measured in the state $\ket{\pm}_p$. Functions which depend only on the $n$ first data $(s_1,\cdots,s_n)$ define a natural filtration ${\cal F}_n$ \cite{probas}, and $p^\pm_n:=\mathbb{E}[\mathbb{I}_{\{s_{n+1}=\pm\}}|{\cal F}_n]={\rm tr}(B_\pm \rho_n B_\pm^\dag)$ are the probabilities for $s_{n+1}=\pm$ conditioned on the value of the internal state at the $n$-th step. A quick and neat way to obtain the scaling limit consists in decomposing the process $\rho_n$ as a sum of a martingale $M_n$ plus a predictable process $O_{n}$. This is called a Doob decomposition \cite{probas}. In the scaling limit the martingale (resp. predictable) contribution converges to the noisy source (resp. the drift) of the SDEs. Eq.(\ref{eq:dilatOQRW}) may be tautologically written as:
\[ \rho_{n+1}= \rho_n^{(+)}\, \mathbb{I}_{\{s_{n+1}=+\}} + \rho_n^{(-)}\, \mathbb{I}_{\{s_{n+1}=-\}} \]
and $x_{n+1}-x_n=\mathbb{I}_{\{s_{n+1}=+\}} -\mathbb{I}_{\{s_{n+1}=-\}}$, with $\rho_n^{(\pm)}:={B_\pm \rho_n B_\pm^\dag}/{p^\pm_n}$. By construction the Doob martingale is $M_n=\sum_{k=1}^n\pi_k$ with $\pi_k:=\rho_k-\mathbb{E}[\rho_k|{\cal F}_{k-1}]$ given by:
\[ 2\pi_k=\big(\rho_k^{(+)}-\rho_k^{(-)}\big)\big(\mathbb{I}_{\{s_{k+1}=+\}} -p^+_k+p^-_k -\mathbb{I}_{\{s_{k+1}=-\}}\big).\]
The predictable process is defined by complementarity $O_{n}:=\rho_n-M_n$. Taking the scaling limit $\epsilon\to0$, $t=n\epsilon$ fixed, is a matter of Taylor expanding $dM_t:=M_{n+1}-M_n$,  $d\rho_t:=\rho_{n+1}-\rho_n$ and $dX_t:= \sqrt{\epsilon}(x_{n+1}-x_n)$. Identifying $\epsilon$ with $dt$, we get $dM_t=D_N(\rho_t)\, dB_t$, and
\begin{eqnarray}
d\rho_t&=&\big( -i[H,\rho_t]+L_N(\rho_t)\big)dt + D_N(\rho_t)\, dB_t, \label{eq:drho}\\
dX_t&=&U_N(\rho_t)\,dt + \, dB_t,\label{eq:dX}
\end{eqnarray}
with $B_t$ a normalized Brownian motion, $D_N(\rho):=N\rho+\rho N^\dag -\rho\,U_N(\rho)$ and $U_N(\rho):={\rm tr}(N\rho+\rho N^\dag)$. Not surprisingly, eq.(\ref{eq:drho}) is of Belavkin's type \cite{barchielli, belavkin}. The drift in eq.(\ref{eq:dX}) is governed by the internal state and this is responsible for the behaviours observed in Figs.\ref{fig:discrete_ballistic}$\&$\ref{fig:discrete_oscillant}. Let us emphasize that this SDE does not contain any jump process (which are nevertheless allowed in general Belavkin equations). As we will see in the following section, jump statistics do not necessarily emerge directly from scaling limits but can be a simple consequence of a non linearity in the SDE.

\section{Bi-stability and ballistic diffusion}
We take $H=\omega_0\, \sigma^2$ and $N=a\, \sigma^3$. Eqs.(\ref{eq:drho},\ref{eq:dX}) are then compatible with reality of the internal density matrix. We parametrize it as $\rho_t=\half(\mathbb{I}+ q_1\sigma^1+ q_3\sigma^3)$ with $q_1^2+q_3^2\leq 1$. Eqs.(\ref{eq:drho},\ref{eq:dX}) then reads:
\begin{eqnarray*}
dq_3&=& 2\omega_0\, q_1\,dt + 2a(1-q_3^2)\, dB_t\\
dq_1&=& -2(\omega_0\, q_3+a^2\, q_1)\,dt - 2a\, q_1q_3\, dB_t
\end{eqnarray*}
One can check again the convergence to pure states that has been shown in the discrete case, let $\Delta_t:=\det\rho_t$. We have $d\Delta_t^{1/2}=\Delta_t^{1/2}[-2a^2 dt+aq_3dB_t]$ with non positive drift so that $\Delta_t^{1/2} $ is a sub-martingale \cite{probas}. It converges exponentially quickly to $0$, so we may describe $\rho_t$ as a pure state, $q_1=\sin\theta$, $q_3=\cos\theta$. The angle $\theta_t$ then satisfies
\begin{equation} \label{eq:dtheta}
d\theta_t=-2(\omega_0+a^2\sin\theta_t\cos\theta_t)dt - 2a\sin\theta_t\, dB_t
\end{equation}
The behavior of $\theta_t$ is quantitatively different depending whether $a^2\gtrless \omega_0$, and this corresponds to the two regimes we mentioned. For $a^2<\omega_0$, $\theta_t$ rotates randomly but regularly enough around the unit circle, so that the internal state $\rho_t$ oscillates almost regularly. For $a^2>\omega_0$, $\theta_t$ is trapped during random periods in the vicinity of $\theta^*_-\simeq0^-$ or $\theta^*_+\simeq\pi^-$. The points $\theta^*_\pm$ are the minima of the effective potential obtained from eq.(\ref{eq:dtheta}) once correctly normalized. Although it fluctuates, $\theta_t$ turns predominantly clockwise (for $\omega_0>0$) around the unit circle, never crossing back $0$ or $\pi$ anticlockwise. 

%\begin{figure}
%\mbox{\includegraphics[width=0.35\textwidth]{ywalka4.png}}
%\caption{\emph{A $y_t$-trajectory for $a^2>\omega_0$, obtained by numerically solving eq.%(\ref{eq:dtheta}) for $\theta_t$, and then changing to the variable $y_t$.}}
%\label{fig:ywalk}
%\end{figure}

\begin{figure}
\mbox{\includegraphics[width=0.46\textwidth]{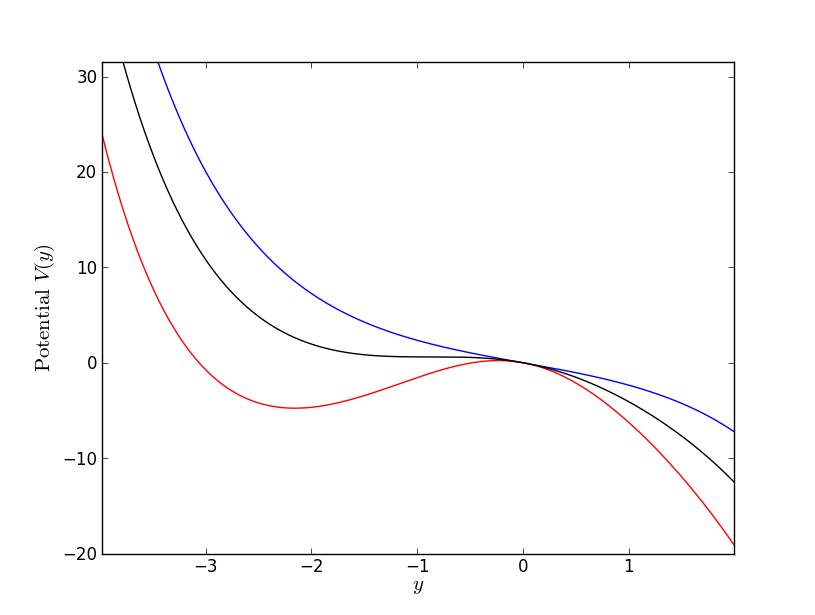}}
\caption{(Color online) Potential $V(y)$: $\omega_0$ is fixed to 1, in red (lowest curve) $a=2$ the potential shows a minimum, in black (middle curve) $a=1$ gives the limiting case, and in blue (highest curve) $a=0$ it has no minimum. }
\label{fig:potential}
\end{figure}

To make this description quantitative, let $y_t:=-\log|\tan\theta_t/2|$. It satisfies a normalized SDE with constant noise source, $dy_t= 2a\, dB_t - V'(y_t) dt$ with potential
\[ V(y)=-2(\pm \omega_0\sinh y +2a^2 \log\cosh y).\]
The above sign is that of $\tan\theta_t/2$, i.e. $+/-$ for $\theta_t$ on the upper/lower half unit circle. What happens in these two sectors is symmetrical, so we concentrate on the upper sector (see Fig. \ref{fig:potential}). The potential shape is that of a cubic like function but it is exponentially large for large $|y|$, i.e. $V(y)\simeq -\omega_0\, {\rm sign}(y) e^{|y|}$. It possesses a minimum and a maximum for $a^2>\omega_0$, and none if $a^2<\omega_0$ (see Fig. \ref{fig:potential}. The minimum is at $y^*_+\simeq -2a^2/\omega_0$ for large $a$, i.e. $\tan\theta^*_+\simeq e^{-y^*_+}$ so that $\theta^*_+$ is close to $\pi^-$, with $V_{\rm min}\simeq -4a^2\log a^2/\omega_0$ and $V_{\rm max}\simeq 0$. When $\theta_t$ enters the upper sector, it does it from $\pi$. For $y_t$ this corresponds to $-\infty$, so that $y_t$ experiences an exponentially steep down ramp that it can never climb back, and this means that $\theta_t$ never escapes the upper sector from $\pi$ but only from $0$. Going down on the 
ramp, $y_t$ reaches the potential 
minimum and spends time fluctuating around there, and this means that $\theta_t$ fluctuates around $\theta_+^*$. At a random time $\tau_{\rm flip}$, large fluctuations allow $y_t$ to cross the energy barrier in a Kramer's like process. Once this has happened, $y_t$ is again on a steep ramp that it steps down to $+\infty$, and this translates to $\theta_t$ moving toward $0^+$ and crossing it irreversibly towards the lower sector. The process then starts on the lower half circle and repeats itself. We estimate the mean flip time as $\mathbb{E}[\tau_{\rm flip}]\simeq e^{\Delta V/4a^2}\simeq a^2/\omega_0^2$ by Kramer's rule, and a more precise study allows us determine the probability distribution of $\tau_{\rm flip}$. 

The internal state drives the system position via eq.(\ref{eq:dX}) which reads $dX_t= 2a\cos\theta_t\, dt +dB_t$. The slopes of the seesaw profiles of $X_t$ are $2a\cos\theta^*_\pm\simeq \mp 2a$. Fluctuations are negligible for large $a$ but the noise is instrumental for tilting from one slope to the other via Kramer's transitions. The mean system density matrix heuristically introduced above is rigorously defined by
\begin{eqnarray} \label{eq:meanrho}
 \hat{\rho_t} := \int {\hskip -0.1 truecm} dx\, \rho(x,t)\otimes \ket{x}_o\bra{x}:=\mathbb{E}[\,\rho_t\otimes \ket{X_t}_o\bra{X_t}\,].
 \end{eqnarray}
Routine applications of stochastic It\^o calculus \cite{probas} show that the SDEs (\ref{eq:drho},\ref{eq:dX}) imply eq.(\ref{eq:FPrho}) for $\rho(x,t)$. Eq.(\ref{eq:FPrho}) is of Lindblad form on ${\cal H}_c\otimes \mathbb{L}^2(\mathbb{R})$. Indeed $\hat{\rho_t}$ verifies:
\begin{equation}
\begin{split}
 \partial_t \hat{\rho_t} = -i[H,\hat{\rho_t}]+L_N(\hat{\rho_t})-\half [P,[P,\hat{\rho_t}]] \\
 - i \left( N [P,\hat{\rho_t}]+[P,\hat{\rho_t}]N^\dagger\right) 
 \end{split}
\end{equation}
with $P=-i\partial_x$ the momentum operator. This formally shows that $\hat{\rho_t}$ defines a completely positive map on ${\cal H}_c\otimes \mathbb{L}^2(\mathbb{R})$.
It may be used to check that $X_t/\sqrt{t}$ becomes Gaussian at large time, in a way compatible with the central limit theorem of \cite{attal2,konno}, and $\mathbb{E}[X_t^2]\simeq D_{\rm eff}\, t$ with effective diffusion constant $D_{\rm eff}= 1 + 4a^4/\omega_0^2$ \cite{BBT}. The factor $1$ comes from the bare diffusion constant \cite{scaling} while the second term, which dominates for large $a$, is induced by the ballistic seesaws. 

\section{Conclusion}
The transition from usual diffusion to ballistically induced diffusion is an echo of the internal gyroscope behaviors. In the ballistic regime the internal state switches randomly between two pure states in a way similar to Kramer's transition. Since the system position is slave to the out-put measurements, our results about convergence from a mixed state to pure states and about random flips between them apply to the coupled probe plus internal spin system without considering orbital d.o.f's. Our analysis of the continuous scaling limit leads us to define the open quantum Brownian motion. More details will be given elsewhere \cite{BBT}.
In the ballistic regime the effective diffusion constant is much larger than the bare one, and one may wonder about other scenarios of ballistically induced diffusion providing large effective diffusion constants.

\begin{acknowledgments} 
This work was in part supported by ANR contract ANR-2010-BLANC-0414. 
D.B. thanks J.M. Raimond for discussions, especially on possible experimental scenarios.
\end{acknowledgments}

\appendix

\section{A simple toy model}
In this section, we study an rather simple model of diffusion with large mean free path whose p.d.f. behaves like the OQRW we studied for large $a$. In this simplified example, the absence of the small Brownian fluctuations around the ballistic trajectories will give a clearer understanding of the non Gaussianity previously observed on the p.d.f.s .

\begin{figure}[!htbf]
 \includegraphics[scale=1.0]{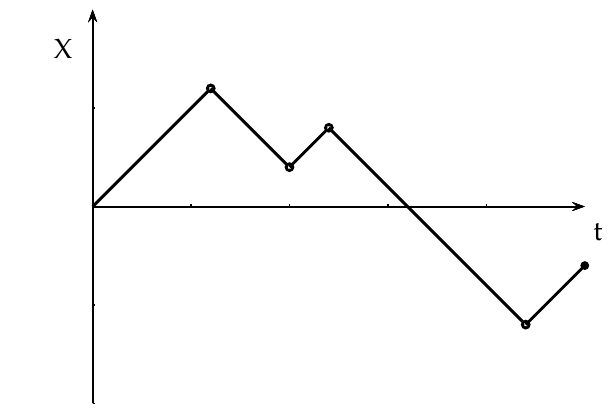}
 \caption{A typical trajectory of the walker on the line}
 \label{fig:toymodel}
\end{figure}

Let us consider a walker continuously going back and forth on the line with a speed $\pm 1$ whose changes are triggered by a counting process of intensity 1 (see Fig. \ref{fig:toymodel}). Such a process can be written rather pompously with a trivial SDE :
\begin{equation}
dX_t=(-1)^{N_t} dt
\end{equation}
Where $N_t$ is just a normalized counting process. We add the initial condition that the walker starts with velocity $+1$. This is just a simplified Brownian trajectory with a large mean free path and we expect a Gaussian probability distribution for large time. We are interested here in the short time behavior of this probability density. As the process is not Markovian in position, one needs to introduce $p_\pm(x,t)$ the probabilities to be in $x$ at time $t$ with a speed $\pm 1$. We collect those 2 probabilities into a vector $\vec{P}$ which can easily be shown to verify the following Fokker-Planck equation:
\begin{equation}
\partial_t \vec{P} + 
\left(\begin{array}[c]{rr}
1 & 0 \\
0 & -1
\end{array}\right)
 \partial_x \vec{P} + 
 \left(\begin{array}[c]{rr}
1 & -1 \\
-1 & 1
\end{array}\right)
\vec{P} = 0
\end{equation}
Notice the similarity with the Fokker-Planck equation verified by the OQRW we previously studied. This equation has no second order term and the large time Gaussian profile only comes from the connection of two transport parts.

\begin{figure}
 \includegraphics[scale=1.0]{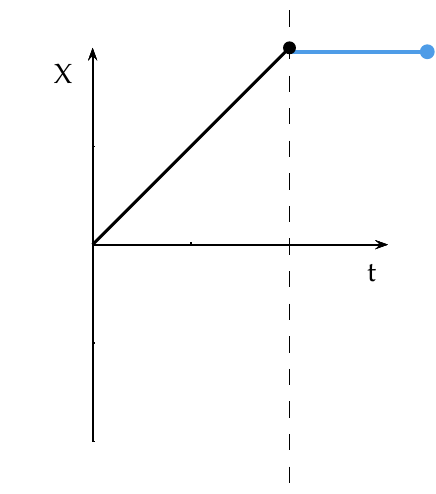}
 \includegraphics[scale=1.0]{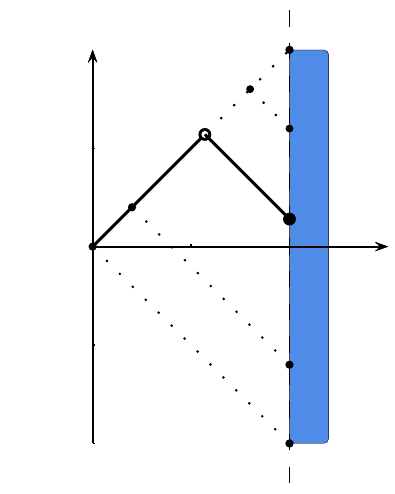}
 \caption{(Color online) Illustration of the expansion in number of flips for the first two terms. On the left, zero flip, only one possible trajectories which gives a Dirac mass. On the right, one flip, one possible trajectory for every end point and thus a uniform distribution.}
 \label{fig:flipexpansion}
\end{figure}

The short time behavior can be understood using an expansion of the probability distribution function $p$ in number of velocity changes or flips (see Fig. \ref{fig:flipexpansion}). 
\begin{equation}
p(x,t)= \sum_{i=0}^{+\infty} P(i \; \mathrm{flips}) P(x,t | i \; \mathrm{flips})
\end{equation}
The first few terms can be computed geometrically and give :
\begin{equation}
p(x,t)=\mathbf{1}_{[-x,x]}(t) \; \left\lbrace \delta(x-t) + \frac{1}{2} + \frac{t}{4}(x+1) + \BigO{t^2}  \right\rbrace e^{-t}
\end{equation}
where $\mathbf{1}_{I}$ is the indicator function of the interval $I$. The Dirac peak corresponds to the situation when no flip occurred. Indeed, in that case the walker goes straight up and all the weight is concentrated on a single point. In the full OQRW model, this straight line is blurred, even for very large $a$, by  small Brownian fluctuations. This noise just changes the p.d.f by a convolution with a sharp Gaussian and the behavior observed remains qualitatively the same.  The constant term corresponds to 1 flip, and the reader can easily get convinced that 1 flip indeed gives rise to a uniform distribution and that it is what explains the plateau in Fig. \ref{fig:pdf}. A similar analysis can be carried on with the linear term corresponding to 2 flips and so on. 
This example shows that puzzling non Gaussianities on p.d.f. can be easily understood when one looks directly at trajectories, which in the case of OQRW are also observable and thus physical.

\begin{references}
\bibitem{Qnoise} In mathematical terms, this defines a dilation of the OQRW quantum map, which is reproduced back when tracing out the probe space. On contrary, quantum trajectories emerge when performing probe measurements and keeping track of their out-puts.
\bibitem{problem} Obviously, the tremendous experimental problem is to act on trapped ions with a few photons and not with a pulse containing a large number of them.
\bibitem{classical_slave} The system position behaves classically if it initially starts localised and if one always measures the probe observable with eigenbasis $\ket{\pm}_p$, as done here. In such case, a quantum  implementation of the orbital d.o.f's is not really needed. However, this is clearly a restriction and one could decide either to initially start with a more general state or to measure other observables.
\bibitem{scaling} Here we make a choice of length and time normalisations, and this fixes the bare diffusion constant $D_0=1$.
\end{references}
\end{document}